\documentclass[aip,reprint]{revtex4-1}

\usepackage{graphicx}

\usepackage{amsmath}
\usepackage{amsthm}
\usepackage{amssymb}
\usepackage{graphpap}
\def\VFT#1{{{\overline{#1}}}}

\def\Real{{\mathbb R}}
\def\Cmpx{{\mathbb C}}

\def\dual#1{{\widetilde{#1}}}

\def\innerprod(#1,#2){{\left<#1\,,\,#2\right>}}
\def\Set#1{{\left\{#1\right\}}}
\def\Re{{\textup{Re}}}
\def\Im{{\textup{Im}}}

\def\qquadtext#1{{\qquad\text{#1}\qquad}}
\def\qquadand{\qquadtext{and}}
\def\quadtext#1{{\quad\text{#1}\quad}}

\def\pfrac#1#2{{\frac{\partial #1}{\partial #2}}}

\def\Cd{{\dot C}}

\def\imu{{\mu}}
\def\inu{{\nu}}
\def\isigma{{\sigma}}

\def\Vx{{\boldsymbol{x}}}

\def\MX{{M_X}}
\def\MY{{M_Y}}

\def\PiY{p_Y}

\def\detg{|\det g|}

\def\rootdetg{\detg^{1/2}}

\def\MD^#1_#2{{\boldsymbol D}^{#1}{}_{#2}}
\def\ML^#1_#2{{\boldsymbol L}^{#1}{}_{#2}}
\def\ID_#1^#2{{\boldsymbol D}_{#1}{}^{#2}}
\def\IL_#1^#2{{\boldsymbol L}_{#1}{}^{#2}}

\def\speciesa{{\text{\tiny$\lfloor\!\alpha\!\rceil$}}}
\def\speciesabig{{\text{$\lfloor\!\alpha\!\rceil$}}}
\def\specieselc{{\text{$\lfloor\!\textup{el}\!\rceil$}}}
\def\speciesion{{\text{$\lfloor\textup{ion}\!\rceil$}}}

\def\kHat{{\hat \kappa}}
\def\FT#1{\widehat{#1}}

\def\Ve{{\boldsymbol e}}
\def\Vd{{\boldsymbol d}}
\def\Vb{{\boldsymbol b}}
\def\Vh{{\boldsymbol h}}

\def\I{{\textup{I}}}
\def\II{{\textup{I\!I}}}
\def\III{{\textup{I\!I\!I}}}

\def\mag{{\cal M}}
\def\pol{{\cal P}}
\def\kappaQM{{\cal Q}_0}

\newsavebox{\CD}
\sbox{\CD}{$i_{\Cd^\speciesa} F_0$}

\begin{document}
\title{Covariant Constitutive Relations, Landau Damping and
  Non-stationary Inhomogeneous Plasmas}
\author{J Gratus and R W Tucker}
\affiliation{Department of Physics, Lancaster University and The
  Cockcroft Institute, Keckwick Lane, Daresbury, WA4 4AD, UK}
\begin{abstract}
Models of covariant linear electromagnetic constitutive relations are formulated that have wide applicability to the computation of susceptibility tensors for dispersive and inhomogeneous  media. A perturbative  framework is used to derive a linear constitutive relation for a  globally neutral plasma enabling  one to describe in this context a generalized Landau damping mechanism for non-stationary inhomogeneous plasma states.

\end{abstract}
\maketitle
Constitutive relations are widely used when describing the behaviour of electromagnetic fields in continuous media. Although their domain of applicability is often determined experimentally, causality and locality play important roles in their theoretical foundations.
Limitations arise due to the inherent
non-linearity contained in the classical equations describing the
coupling between the motion of  individual particles or continuous charge distributions and a self-consistent electromagnetic
field. For small disturbances perturbative linearisation techniques
are available and approximation schemes exist for calculating
effective susceptibility tensors that arise from these constitutive relations. The effective constitutive relations
that result from such schemes often rely for their validity on
assumptions such as material homogeneity and non-relativistic
perturbations about stationary  configurations. This letter
addresses some of the issues that arise when some of these assumptions
are relaxed and the degree to which concepts such as Landau damping
can be generalized for relativistic inhomogeneous plasmas \cite{o1973relativistic}, \cite{beskin1987permittivity}, \cite{caldela1989dispersion}.

The
{\it macroscopic} Maxwell equations can be written as the exterior
system
\begin{align}
d\,F=0,  \qquad\qquad \epsilon_0 d\,\star G=-\star\dual{J}
\label{Maxwells_eqns}
\end{align}
where $F$ is the Maxwell 2-form, $G=\epsilon_0 \,F+ \Pi$ the
excitation 2-form and $J$ the source current vector on a spacetime   $M$, in
terms of the Hodge map $\star$ and metric dual $\dual J$ associated
with the spacetime metric $g$.
In this form certain 4-current sources $-d\star\Pi$  are
 included in $G$ with  the remaining 4-currents denoted by $J$.
A covariant constitutive
model provides functional relations between $G$ (or $\Pi$) and $F$ and
between $J$ and $F$.
Relative to any unit future-pointing
timelike observer vector field $U$ on $M$, the forms $F$ and $G$
define the frame dependent electromagnetic 1-forms $\Ve^U= i_U\,F$,
$\Vb^U = i_U\star F$, $\Vd^U= i_U G$ and $\Vh^U= i_U \star G$ so that
$F = \dual{U}\wedge \Ve^U - \star(\dual{U}\wedge\Vb^U)$ and likewise
for $G$.

If one restricts to causal {\it linear} responses a natural covariant
 constitutive relation is given by the non-local expression
\begin{align}
\Pi[F]_{ab}(x)=\tfrac14 \int_{y\in J^{-}(x)} \!\!\chi_{abcd}(x,y) F_{ed}(y)
dy^{cdef}
\label{Story_CR_index}
\end{align}
where $\chi_{abcd}(x,y)$ is a two-point susceptibility kernel. The
events $x$ and $y$ are given in arbitrary coordinates with summation
over Latin indices from $0$ to $3$ and  $dy^{cdef}=dy^c\wedge dy^d\wedge dy^e\wedge dy^f$.  The
causal structure has been imposed by requiring that
$\chi_{abcd}(x,y)=0$ if $y$ does not lie in the past light cone,
$J^{-}(x)$, of $x$.  This constitutive relation can be used to model
media which are spatially inhomogeneous and temporally non-stationary and
is meaningful in spacetimes containing gravitation.

To facilitate the discussion and the  use of two-point tensors  introduce two copies of
$M$, denoted $\MX$ and $\MY$, with generic points
$x\in\MX$ and $y\in\MY$  coordinated by $(x^0,\ldots,x^3)$ and
$(y^0,\ldots,y^3)$ respectively. The values $\chi_{abcd}(x,y)$
denote the  coordinate components of the $4-$form field $\chi$  over the product manifold
$\MX\times\MY$ in the induced  coordinates   $(x^0,\ldots,x^3,y^0,\ldots,y^3)$:
\begin{align}
\chi=\tfrac14 \chi_{abcd}\,
dx^{a}\wedge dx^{b}\wedge dy^{c}\wedge dy^{d}\,.
\label{Disp_coords}
\end{align}
In terms of $\chi$ and the projection $\PiY:\MX\times\MY\to\MY$,
$\PiY(x,y)=y$ equation (\ref{Story_CR_index}) can be written
\begin{align}
\Pi[F]=\int_{ \MY} \chi\wedge\PiY^\star F\,.
\label{PIF}
\end{align}
The tensor $\chi$ has 36 independent components since $dx^{ab}$ and
$dy^{cd}$ are antisymmetric so
$\chi_{abcd}=-\chi_{bacd}=-\chi_{abdc}$.

A special case of (\ref{Story_CR_index}) arises in  Minkowski spacetime.
Being parallelizable it admits a
family of {\it translation maps} $A_z: M\to M, x\mapsto A_z(x)=x+z$
for all events $x$ in $M$. This induces the translation maps
$B_z:\MX\times\MY\to\MX\times\MY$,
$B_z(x,y)=(x+z,y+z)$.
Imposing this translational symmetry on $\chi$,
i.e. $B_z^\star\chi=\chi$,  equation (\ref{Story_CR_index}) can be written
\begin{align}
\Pi_\I[F]_{ab}(x)=\int_{{y\in J^-(x)}} X_{abcd}(x-y) F_{ef}(y) dy^{cdef}
\label{hom}
\end{align}
where $X_{abcd}(x-y)=\chi_{abcd}(x,y)=\chi_{abcd}(x+z,y+z)$ for any
$z$.
Since  $\Pi_\I[A_z^\star F]=A_z^\star\Pi_\I[F]$
this  relation describes  spatially homogeneous and temporally
stationary media  and remains
non-local in both space and time. Such a medium  exhibits both {\it spatial and temporal dispersion  }  as follows:
Minkowski spacetime admits preferred  {\em global Lorentzian} coordinate systems, $(x^0,x^1,x^2,x^3)$, with associated  cobases
$(dx^0,dx^1,dx^2,dx^3)$ in which the components of the metric are
diag$(-1,1,1,1)$.  If one defines for any scalar $\phi$
the Fourier transform
\begin{align*}
\FT{\phi}(k)=
\int_M e^{-ik\cdot x}\,\phi(x)\,dx^{0123}
\end{align*}
with respect to such a coordinate system
then $\widehat{\Pi_\I[F]}_{ab}(k)=\FT{X}_{abcd}(k)\FT{F}_{ef}(k)
\epsilon^{cdef}$ in terms of the standard constant alternating symbol $\epsilon^{cdef}$. Such a relation can give rise to dispersion in media.

If the bulk 4-velocity of a non-accelerating medium is $V$, with constant components  in the above coordinate system,  i.e. $\nabla V=0  $,  a
particular model for $X$ in (\ref{hom}) is given by
\begin{equation}
\begin{aligned}
\Pi_{I_{a}}[F]_{ab}(x)&=
\int_{{y\in J^-(x)}} \pol(x-y) (i_VF\wedge\dual{V})_{ab}(y) dy^{0123}
\\&\hspace{-3em}-
\star_X\int_{{y\in J^-(x)}} \mag(x-y)  (i_V\star_Y
F\wedge\dual{V})_{ab}(y)
dy^{0123}
\end{aligned}
\label{hom_xizeta}
\end{equation}
where  $\pol$ and $\mag$ are  polarization and magnetization susceptibility scalars respectively.   The Fourier transform of (\ref{hom_xizeta}) yields  the simple constitutive relations
for a spatially and temporally dispersive homogeneous isotropic medium:
$\FT{\Vd^V_a}{}(k)=(\epsilon_0+\FT{\pol}(k))\FT{\Ve^V_a}{}(k)$ and
$\FT{\Vh^V_a}{}(k)=(\mu_0^{-1}+\FT{\mag}(k))\FT{\Vb^V_a}{}(k)$.
If $V$ is not inertial ($\nabla V \neq 0$)  then (\ref{hom_xizeta}) is not a special case of (\ref{hom}) and its Fourier transform, although local in $k$, is not of the form above.

For media that lack spatial dispersion the history of the medium may give rise to {\it temporal dispersion} alone. This can be expressed geometrically in terms of  tensor  transport along the integral curves  $C_\Vx:\Real\to M$ of the 4-velocity field  $V$ of the medium. If these curves are each parameterized by proper time $\tau$ let
$\Phi_{\tau}^{\hat\tau}(\Vx)$ be a map
that transports tensors at $C_{\Vx}(\tau)$ to tensors at
$C_{\Vx}(\hat\tau)$ along each integral curve of $V$. Natural
choices of transport maps include Lie, parallel (with respect to some
spacetime connection $\nabla$) and Fermi-Walker transport. Different
choices of $\Phi_{\tau}^{\hat\tau}(\Vx)$ correspond to different
electromagnetic responses of the medium to the disposition of the
integral curves of $V$ in the spacetime history of the medium. If $Y(z)$
denotes  a tensor field mapping 2-forms  at $z$ to 2-forms at $z$, a constitutive relation for  a spatially  inhomogeneous medium  may be written
\begin{align}
&\Pi_{{\II}}[F]{(C_{\Vx}(\tau))}
=
\int_{-\infty}^\tau\hspace{-.7em}
Y^\Phi(\tau,\hat\tau,\Vx)
\Big(\Phi_{\hat\tau}^\tau(\Vx)\big(F{(C_{\Vx}(\hat\tau))}\big)\Big)
d\hat\tau
\label{temphom}
\end{align}
where $
Y^\Phi(\tau,\hat\tau,\Vx)=
\Phi_{\tau-\hat\tau}^\tau(\Vx)\big(Y{(C_{\Vx}(\tau-\hat\tau))}\big)
$ is a tensor at $C_{\Vx}(\tau)$. This is another special case of (\ref{Story_CR_index}).
Since $$ \Pi_{{\II}}[ F^{\Phi}]\big( C_{\Vx}(\hat\tau)   \big) =\Phi^{\hat \tau}_\tau(\Vx)\Big(\Pi_{{\II}}[F]\big( C_{\Vx}(\tau)  \big)   \Big)   $$
where $ F^{\Phi}( C_{\Vx}(\hat\tau)   )  =  \Phi^{\hat \tau}_\tau(\Vx)\big(F( C_{\Vx}(\tau)  )   \big)  $ the medium is said to be stationary with respect to the transport map and hence $V$ and (\ref{temphom}) is valid in any spacetime. This generalizes the notion of a {\it temporally stationary} medium in a spacetime with timelike Killing vectors.
 The temporal dispersive properties of the medium are best defined with respect to a modified Fourier transform that remains valid in  a general spacetime and employs the transport map along the family of curves describing the history of the medium.
For  any tensor field $\alpha$ and  curve $C_\Vx$ define, at the event $C_\Vx(0)$, the tensor:
\begin{align*}
\VFT{\alpha}{(\omega,\Vx)}
=
\int_{-\infty}^\infty e^{-i\omega\tau}\Phi_{\tau}^0(\Vx)
\big(\alpha{(C_\Vx(\tau))}\big) d\tau\,.
\end{align*}
Then then constitutive relation
$\VFT{\Pi_{\II}[F]}{(\omega,\Vx)}=\VFT{Y}{(\omega,\Vx)}
\big(\VFT{F}{(\omega,\Vx)}\big)$ describes an anisotropic, spatially
inhomogeneous but temporally dispersive medium.

If $V$ is geodesic (i.e. $\nabla_V\,V=0$) and
$\Phi_{\hat\tau}^\tau(\Vx)$ describes {\it parallel} transport then
a particular model for $Y$ in  (\ref{temphom}) is given by
\begin{equation}
\begin{aligned}
&{\Pi}_{\II_a}[F]{(C_{\Vx}(\tau))}
\!=\!
\\ &\
\int_{-\infty}^\tau\hspace{-1.em}
\pol(\tau\!-\!\hat\tau,\Vx)
\Big(\Phi_{\hat\tau}^\tau(\Vx)(i_V F\!\wedge\!\dual{V}){(C_{\Vx}(\hat\tau))}\Big)
d\hat\tau
\\&\
-\star
\int_{-\infty}^\tau\hspace{-1.em}
\mag(\tau-\hat\tau,\Vx)
\Big(\Phi_{\hat\tau}^\tau(\Vx)(i_V\!\star\! F\!\wedge\!\dual{V}){(C_{\Vx}(\hat\tau))}\Big)
d\hat\tau
\end{aligned}
\label{new}
\end{equation}
where $\pol$ and $\mag$ are spatially inhomogeneous polarization and magnetization susceptibility scalars respectively.  This describes
 non-magneto-electric, spatially  inhomogeneous  but temporally dispersive media with constitutive relations
$\VFT{\Vd^V}{}(\omega,\Vx)=
(\epsilon_0+\VFT{\pol}(\omega,\Vx))\VFT{\Ve^V}{}(\omega,\Vx)$
and
$\VFT{\Vh^V}{}(\omega,\Vx)=
(\mu_0^{-1}+\VFT{\mag}(\omega,\Vx))\VFT{\Vb^V}{}(\omega,\Vx)$.

The historic covariant constitutive relations proposed by Minkowski belong to a class of
 {\it local linear} relations on spacetime of
the form
\begin{align}
\Pi_\III[F](x)={\cal Z}(x)\big(F(x)\big)
\label{pointwise}
\end{align}
where  the tensor ${\cal Z}(x)$ maps
2-forms at events $x$ to 2-forms at $x$. For a medium with bulk 4-velocity field $V$ this tensor may be chosen
so that
$\Vd^V(x)=(\epsilon_0+\pol(x))\Ve^V(x)$ and
$\Vh^V(x)=(\mu_0^{-1}+\mag(x))\Vb^V(x)$ valid for inhomogeneous and non-stationary isotropic media in an arbitrary spacetime.
However in a Minkowski background their Fourier transforms for
non-constant $\pol$ and $\mag$ yield non-local constitutive
relations among the Fourier components of the electromagnetic fields
so do not describe normal dispersive continua.

The different constitutive relations (\ref{hom}), (\ref{hom_xizeta}),
(\ref{temphom}), (\ref{new}) and (\ref{pointwise}) above are applicable to the
phenomenological description of {\it linear}  media that exhibit temporal
and/or spatial dispersion but  rely on
some knowledge of the electromagnetic response of systems in either
inertial or co-moving reference frames.
However such relations do not encompass the effective susceptibility that arises when one applies a perturbative analysis to processes involving non-stationary, spatially inhomogeneous plasmas  \cite{vladimirov1998covariant,melrose1973covariant,lamalle2002kineticI,lamalle2002kineticIII}.
Such processes are not
uncommon in astrophysical applications or in regimes where
instabilities arise
\cite{weibel1959spontaneously,schaefer-rolffs:012107,yoon:1336}
from inhomogeneities such as those in laser-plasma systems.  In these situations the
{\it microscopic} Maxwell system for a neutral plasma, $d\,F=0,\,
\epsilon_0\,d\,\star F=-\star\dual{J}$, can be coupled with equations for the
charged sources $J$ in order to effect a linearisation.  Since
$d\,\star\dual{J}=0$ it is convenient to write $\star\dual{J}= -d\star\Pi$ and
$\epsilon_0\,F=G-\Pi$, to cast the system into the form used above for
neutral polarizable media. Writing $\Pi$ as a functional of $F$ in the form  (\ref{Story_CR_index}), enables one to derive
an effective  constitutive relation for a globally neutral plasma.


\section{The covariant susceptibility kernel for a non-stationary,
spatially  inhomogeneous plasma}
Consider a macroscopically neutral plasma composed of particles
labeled by the species index $\speciesabig$, mass $m^{\speciesa}$ and
charge $q^{\speciesa}$, described dynamically by the
coupled relativistic (collisionless) Maxwell-Vlasov equations for $F$
and the one-particle probability distributions $f^\speciesa(x,v)$ in
an arbitrary (background) gravitational field:
\begin{align}
W^\speciesa(f^\speciesa)=0\,,
\label{Plasma_MV_f}
\end{align}
\begin{align}
dF=0\qquadand \epsilon_0 d\star F=-\star \dual{J}
\label{Plasma_MV_Ff}
\end{align}
with the total current 1-form ${\tilde J(x)}= g_{ab}J^b(x)\,dx^a$ given by
\begin{align}
J^b(x)=
-\sum_\speciesa
q^\speciesa\int \frac{v^b\rootdetg}{v_0}
f^\speciesa(x,v) dv^{123}
\label{Plasma_Jf_sp}
\end{align}
and Liouville vector field $W^\speciesa$
\begin{align}
&W^\speciesa{(x,v)}=
\label{Plasma_MV_W}
\\&\quad v^a\pfrac{}{x^a}+
\big(-\Gamma^\inu{}_{ef}(x)v^e v^f +
\frac{q^\speciesa}{m^\speciesa}F_{ef}(x) g^{\inu e} v^f\big)\pfrac{}{v^\inu}
\notag
\end{align}
with summation over Greek indices from 1 to 3.
 The function $v^0(x,v^1,v^2,v^3)$  is a solution of  $v^a v^b g_{ab}(x)=-1$.

A linear constitutive relation arises from a perturbation of this
system about a background (zeroth-order) spatially inhomogeneous  and temporally non-stationary solution, $F_{0}(x), f^\speciesa_0(x,v)$.
The standard perturbation
expansion  about such a solution yields a linear system for $f_1$ and $F_1$ in terms of $f_0$ and $F_0$ that can be solved in principle by
the method of characteristics \cite{brambilla1999high}.
This yields a first order system of integro-differential equations for $F_1$:
\begin{align}
d\,F_1=0,\qquad
d\star\,F_1= - d\star \Pi_1[F_1]\label{dstarF1}
\end{align}
where $\Pi_1$ is  a linear functional depending on $f_0$ and the   solutions  to the zeroth-order Lorentz force equations
\begin{align}
\nabla_{\Cd^\speciesa}\Cd^\speciesa=
\frac{q^\speciesa}{m^\speciesa}\dual{\mbox{\usebox{\CD}}}
\label{Plasma_Lorentz_Force}
\end{align}
after elimination of $f_1$. This gives rise to a class of solutions $\chi_1$ to (\ref{PIF}) where $\Pi, \chi$ are replaced by $\Pi_1,\chi_1$ respectively.

If one employs standard
inertial coordinates in a Minkowski spacetime and a
zeroth-order electromagnetic field $F_0=0$, in zeroth-order
all particles move along straight line time-like geodesics
independent of $\speciesabig$ and
\begin{align*}
\chi^{\speciesa}_1{(x,y)}
=&
\frac{q^{\speciesa2}}{m^\speciesa}
\frac{ f_0^\speciesa(y,\hat{u})}{4 \hat{u}_0\hat{\tau}^2} g^{\imu c} \hat{u}^b
\epsilon_{cbih}
\\&\times
\big(2dx_{0\mu}+
\epsilon^{d\sigma jk}
\epsilon_{\imu\inu\isigma}
\hat{u}^\inu \hat{u}_d dx_{jk}
\big)
\wedge dy^{ih}
\end{align*}
where
$\hat{\tau}(x,y)=(-g(x-y,x-y))^{1/2}$ and
$\hat{u}(x,y)={(x-y)}/{\hat{\tau}(x,y)}$
which is manifestly not  a function of $x-y$ alone.

\section{Langmuir modes for a neutral non-stationary, spatially inhomogeneous plasma in Minkowski spacetime}

 The general solution to the second equation in (\ref{dstarF1})
 is given by $\epsilon_0\star
F_1=-\star\Pi_1[F_1]+d \beta$ where $\beta$ is an arbitrary 1-form.
We define the generalized Langmuir sector to contain  particular solutions satisfying
\begin{align}
\epsilon_0 F_1=-\Pi_1[F_1]\,.
\label{Plasma_disp_restrict}
\end{align}
This then reduces to the standard perturbative solution describing longitudinal plasma oscillations about a {\it stationary}  $f_0$  in  a {\it spatially homogeneous}  plasma.
More generally consider the situation where  $F_0=0$ and planar inhomogeneities arise
from the following zeroth-order non-stationary spatially inhomogeneous solution to the Maxwell-Vlasov system:
(\ref{Plasma_MV_f}-\ref{Plasma_MV_W})
\begin{equation}
\begin{aligned}
f_0^\specieselc(t,\xi,x^2,x^3,u,v^2,v^3)
&=
f_0^\speciesion(t,\xi,x^2,x^3,u,v^2,v^3)
\\&\hspace{-3em}=
h\Big(\xi-\frac{u \xi}{({1+u^2})^{1/2}},u\Big) \delta(v^2) \delta(v^3)\,.
\end{aligned}
\label{eg_dist}
\end{equation}
If
\begin{align*}
h(\xi,u) = n^\speciesion(\xi) A^\speciesion(\xi)
\exp\Big( - \frac{m^\speciesion({1+u^2})^{1/2}}{k_B T^\speciesion(\xi)} \Big)
\end{align*}
where $A^\speciesion(\xi)$ normalizes (\ref{eg_dist}),
then $f^\speciesion$ initially at $t=0$ represents a distribution of
ions where, at each spatial point $\xi$, the velocities belong to the   1-dimensional
Maxwell-J\"uttner distribution, but where the temperature
$T^\speciesion(\xi)$ and the number density of ions $n^\speciesion(\xi)$,
depend on position. It follows from (\ref{eg_dist}) that
$f^\specieselc$ also initially represents a position dependent
Maxwell-J\"uttner distribution where
$n^\specieselc(\xi)=n^\speciesion(\xi)$ and
$T^\specieselc(\xi)=T^\speciesion(\xi)m^\specieselc/m^\speciesion$. After
the initial moment, the ions and electrons  drift according to
(\ref{eg_dist}) and velocities do not remain in the
Maxwell-J\"uttner distributions.
Alternatively (\ref{eg_dist}) might describe a plasma composed of particles and
anti-particles.

In the theory of a homogeneous plasma one has the solution
$F_1=dt\wedge d\xi \,\FT{E}(\omega,\kappa)\, e^{-i\omega t + i \kappa
  \xi}$ provided $\omega$ and $\kappa$ satisfy a transcendental
dispersion relation. This relation contains an integral that is
potentially singular. The Landau prescription circumvents this
singularity by complexifying $\omega$  and defining an
analytic continuation for the integral in the complex $\omega$ plane.
In an inhomogeneous plasma there is no such time harmonic solution or
associated algebraic dispersion relation between $\omega$ and
$\kappa$.  We therefore propose solving (\ref{Plasma_disp_restrict})
with a longitudinal field $F_1$ represented as
\begin{align*}
F_1{(t,\xi)}=dt\wedge d\xi \int_{\omega=-\infty}^\infty d\omega
\int_{\kappa=-\infty}^\infty d \kappa\
\FT{E}(\omega,\kappa) e^{-i\omega t + i \kappa \xi}\,.
\end{align*}
In this case the Landau dispersion relation is replaced by an integral equation for $ \FT{E}(\omega,\kappa)  $. This equation contains a double integral that requires analytic continuation in the complex $\omega$ plane for its  definition. If one restricts to modes with real $\kappa$ there are now two singular branch points in the  $\omega$ plane  at $\omega=\pm\vert \kappa \vert$.
 To define an analytic continuation from the upper-half  $\omega$ plane the associated cuts are disposed along the half-lines
$\Set{\omega=-|\kappa|-i\lambda,\lambda>0}$ and
$\Set{\omega=|\kappa|-i\lambda,\lambda>0}$ (see figure).

\begin{center}
\setlength{\unitlength}{0.03\textwidth}
\begin{picture}(10,8)
\put(0,0){\includegraphics[height=8\unitlength]{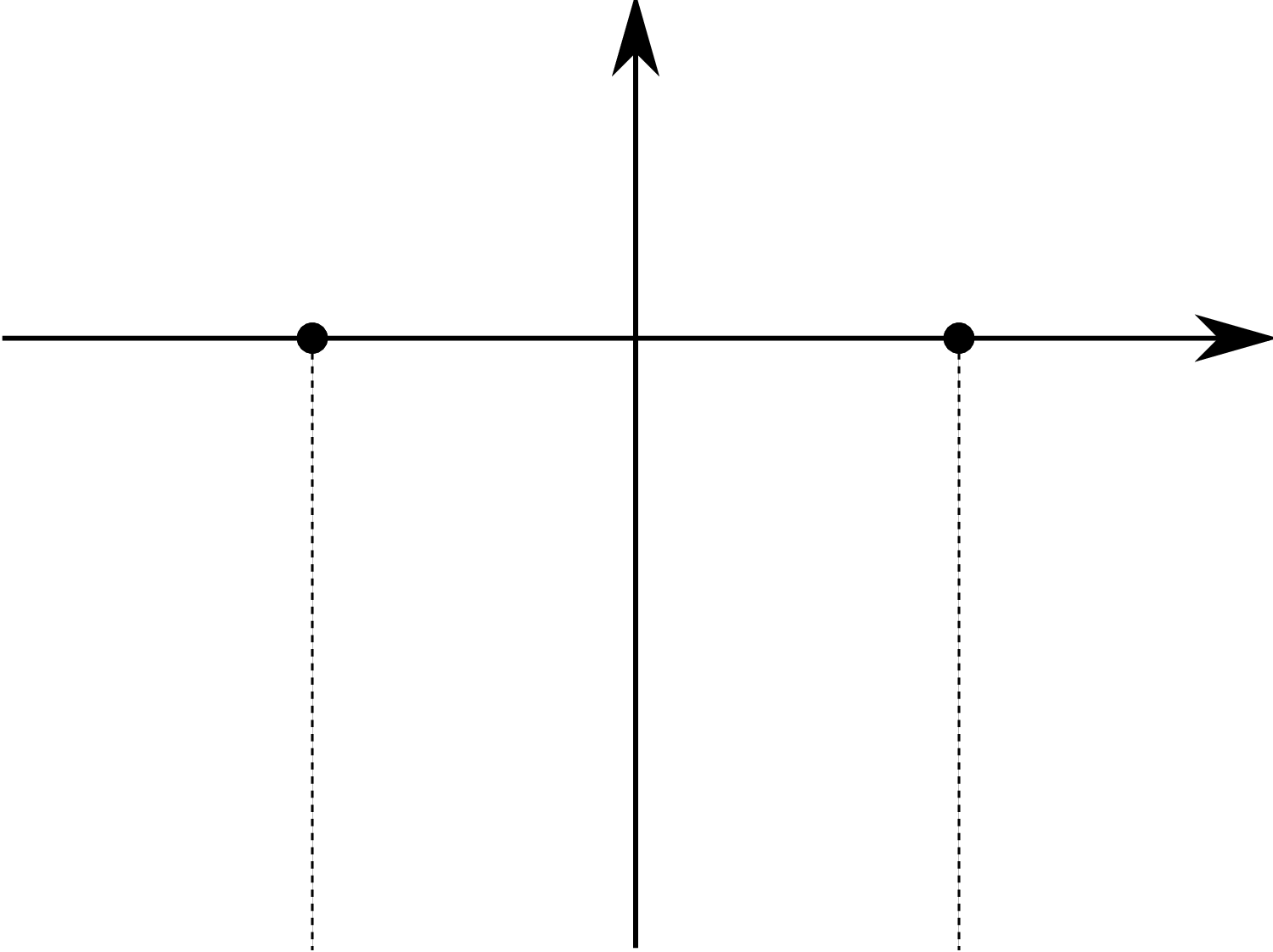}}
\put(8,7){$\omega$ - $\Cmpx$-plane}
\put(5,7){\makebox(0,0)[r]{$\Im(\omega)$}}
\put(11,5){\makebox(0,0)[tr]{$\Re(\omega)$}}
\put(8,5.2){\makebox(0,0)[b]{$|\kappa|$}}
\put(2.5,5.2){\makebox(0,0)[b]{$-|\kappa|$}}
\put(2.1,1){\rotatebox{90}{\small branch cut}}
\put(8.2,1){\rotatebox{90}{\small branch cut}}
\put(5.2,5.1){\makebox(0,0)[ct]{$({\kappa^2-\omega^2})^{1/2}>0$}}
\end{picture}
\end{center}
The integral   equation for
$\FT{E}(\omega,\kappa)$  is now given by
\begin{align*}
\FT{E}(\omega,\kappa)
&=
\frac{\kappaQM}{2\pi}
\int_{-\infty}^\infty
P(\omega,\kappa,\kHat)\,d\kHat
\\&
\hspace{4em}\quadtext{if} \Im(\omega) > 0
\quadtext{or} |\Re(\omega)|>|\kappa|\,,
\\
\FT{E}(\omega,\kappa)
&=
\frac{\kappaQM}{2\pi}
\int_{-\infty}^\infty \hspace{-1em}
P(\omega,\kappa,\kHat)\,d\kHat
-i\kappaQM
\int_{-\infty}^{\infty} \hspace{-1em}
 R(\omega,\kappa,\kHat)\,d\kHat
\\&
\hspace{4em}\quadtext{if} \Im(\omega) < 0\quadtext{and}|\Re(\omega)|\le|\kappa|
\,,
\\
\FT{E}(\omega,\kappa)
&=
\frac{\kappaQM}{2\pi}
\int_{-\infty}^\infty \hspace{-1em}
P(\omega,\kappa,\kHat)\,d\kHat
-\frac{i\kappaQM}{2}
\int_{-\infty}^{\infty}\hspace{-1em}
R(\omega,\kappa,\kHat)\,d\kHat
\\&
\hspace{4em}\quadtext{if} \Im(\omega) = 0\quadtext{and}|\Re(\omega)|<|\kappa|
\end{align*}
where $\kappaQM=\sum_\speciesa q^{\speciesa 2}/\epsilon_0 m^\speciesa$,
the principal  value integral is given by
\begin{align*}
\lefteqn{P(\omega,\kappa,\kHat)=}&
\\&
\int_{-|\kappa-\kHat|}^{|\kappa-\kHat|}
\frac{\FT{E}(\omega\!+\!\kappa',\kHat)(\kappa\!-\!\kHat)}
{(\omega \kappa - \omega\kHat  -\kappa' \kappa)^2}
\FT{h}\bigg(\!\kappa\!-\!\kHat,\frac{\kappa'}
{\big({(\kappa\!-\!\kHat)^2\!-\!\kappa'^2}\big)^{1/2}}\bigg)
\, d\kappa'
\end{align*}
 and the residue
by
\begin{align*}
R(\omega,\kappa,\kHat)
&=
\frac{|\kappa-\kHat|}{\kappa |\kappa|}
\pfrac{\FT{E}}{\omega}\Big(\frac{\omega\kHat}{\kappa},\kHat\Big)
\FT{h}\Big(\kappa-\kHat,\frac{s_\kappa s_{\kappa-\kHat} \omega}
{({\kappa^2-\omega^2})^{1/2}}\Big)
\\& \hspace{-1em}
-
\frac{\kappa}{(\kappa^2-\omega^2)^{3/2}}
\FT{E}\Big(\frac{\omega\kHat}{\kappa},\kHat\Big)
\pfrac{\FT{h}}{u}
\Big(\kappa-\kHat,\frac{s_\kappa s_{\kappa-\kHat} \omega}
{({\kappa^2-\omega^2})^{1/2}}\Big)\,.
\\
\end{align*}
Here $s_\kappa=\kappa/|\kappa|$ and $\FT{h}(\kappa,u) =
\int_{s=-\infty}^\infty e^{-i \kappa s} h(s,u) ds$. The square root
$({\kappa^2-\omega^2})^{1/2}$ is defined so that for $\omega\in\Real$,
$|\omega|<|\kappa|$ and with the branch cuts given in the figure then
$({\kappa^2-\omega^2})^{1/2}>0$.

For real $\kappa$ these integral equations can be analyzed numerically
\cite{petri2007numerical} in the different domains in the $\omega$
plane and wave instability is associated with solutions for which
Im$(\omega) >0 $. Although the nature of Landau damping (Im$(\omega) <
0$) in the presence of inhomogeneities is clearly more complicated
that analogous damping in homogeneous plasmas, the results above
indicate how the mechanism depends on the nature of the initial state
and analytic continuation in the complex $\omega$ plane.

\bibliographystyle{unsrt}


\end{document}